\documentclass[footinbib,a4paper,aps,prb,amsmath,amssymb,showpacs,10pt,onecolumn,floatfix]{revtex4}
\usepackage[latin1]{inputenc}
\usepackage{amsmath}
\usepackage{amsfonts}
\usepackage{amssymb}
\usepackage{graphics}
\usepackage{graphicx}
\usepackage{listings}
\begin{document}
\title{Nonanalytic quantum oscillator image of complete replica symmetry breaking}
\author{R. Oppermann, H. Schenck}
%
\begin{abstract}
We describe the effect of replica symmetry breaking in the field distribution function $P(h)$ of the $T=0$ SK-model as the difference between a split Gaussian and the first excited state $\psi_1$ of a weakly anharmonic oscillator with nonanalytic shift by means of the analogy $P(h)\leftrightarrow |\psi_{_1}(x)|$.

New numerical calculations of the leading $100$ orders of replica symmetry breaking (RSB) were performed in order to obtain $P(h)$, employing the exact mapping between density of states $\rho(E)$ of the fermionic SK-model and $P(h)$ of the standard model, as derived by Perez-Castillo and Sherrington.
Fast convergence towards a fixed point function $\rho(E)$ for infinite steps of RSB is observed.
A surprisingly small number of harmonic oscillator wave-functions suffices to represent this fixed point function. This allows to determine an anharmonic potential $V(x)$ with nonanalytic shift, whose first excited state represents $\rho(E)$ and hence $P(h)$.
The harmonic potential with unconventional shift
$V_2(x)\sim (|x|-x_0)^2=(x-x_0\,sign(x))^2$ yields already a very good approximation,
since anharmonic couplings of $V(x)-V_2(x)\sim |x|^{m}, m>2,$ decay rapidly with increasing $m$.

We compare the pseudogap-forming effect of replica symmetry breaking, hosted by the
fermionic SK-model, with the analogous effect in the Coulomb glass as designed by Davies-Lee-Rice and described by M\"uller-Pankov.

%
\end{abstract}
\maketitle
\section{Introduction}
The Sherrington-Kirkpatrick (SK-) model \cite{SK} has been an extraordinary stimulating source for analytical and numerical studies in the physics of frustrated magnetism and in numerous interdisciplinary applications. Many relevant properties of the SK-model have been explored since its discovery \cite{mezard,parisi,thomsenthorpe,talagrand,sommers}.

Near the zero temperature limit and at $T=0$, the model displays rather simple features but also analytically still unresolved behavior, and therefore remains challenging in our point of view.


Along a particular line of theoretical analysis, on which the present article evolves, we searched for explicit analytical solutions deep in the ordered phase and particularly at zero temperature. The limit of zero temperature was recently described to accommodate two critical points related to the criticality of the hierarchical scheme of replica symmetry breaking (RSB) hosted by the SK-model \cite{prl2005,prl2007,mjs-ro-pre,ro-mjs-pre}.
In order to reveal this type of criticality, the finite-step discrete-RSB was utilized.
Extreme high order calculations up to $200$ orders and Pad\'e-approximants allowed to analyze the limit towards infinite breaking analyzed by means of high-precision numerical work, renormalization group ideas, and scaling theory in addition.
\cite{prl2005,prl2007,mjs-ro-pre,ro-mjs-pre}. The scaling theory allowed to resolve the puzzle of noncommuting zero temperature limit and infinite-step limit of replica symmetry breaking \cite{ro-mjs-pre}.



In this article we present new numerical calculations which led us to a surprisingly simple modeling of the distribution of internal magnetic fields.

Pioneering work on this local magnetic field distribution by Thomsen et al \cite{thomsenthorpe} presented various analytical limits and numerical studies (TAP theory, simulations etcetera) for vector spin glasses (incl. SK of course) and for the relevant temperature range from zero to above the glass transition.

By virtue of an exact mapping as derived by Perez-Castillo and Sherrington \cite{perez},
we can, for our convenience, evaluate the spectral function (or density of states) for the fermionic version of the SK-model (see below for definition) and directly employ its functional form in one-to-one correspondence as the field distribution of the standard SK-model.

In the following section we recall a few necessary details of the fermionic SK-model (sometimes called "ISG$_f$" or "fSK") and of the fermionized SK-model as an exact representation of the standard SK-model. The formalism on how to deal with the Parisi scheme in the Grassmann field theories of these models has been presented in numerous earlier papers at length.

Our main goal here is to report an astonishingly simple representation of the internal magnetic field distribution by the functional form of a single eigenstate of a weakly anharmonic quantum oscillator.

This idea was guided by the known simple split-Gaussian replica-symmetric solution for the density of states $\rho(E)$ at $T=0$.
The broad energy gap of this lowest approximation is filled by replica symmetry breaking. The resulting pseudogap reflects the criticality of RSB at $T=0$. Its power law must be compared with the Efros-Shlovskii pseudogap, and indeed we will add below remarks on the Coulomb glass pseudogap as derived by M\"uller and Pankov \cite{mp}.
Their intricate selfconsistent scheme \cite{mp} is used to unravel the special role of the Coulomb interaction and of RSB for the pseudogap-creation at very low temperatures.

The representation of the spectral density by means of 1D quantum oscillator wavefunctions was also motivated by the advantage that one can integrate analytically the Lehmann representation of the single fermion Greens function (term by term) for the fermionic SK-model.

The paper is organized as follows:

The main results on the functional mapping between field-distribution $P(h)$, density of states $\rho(E)$ and the first excited state of an unconventional quantum oscillator with nonanalytic shift and very weak anharmonicity, is addressed in all chapters. The numerical data, on which the partially-analytical proposal is based, are presented and analyzed in section 4. The analytical formulas, which were evaluated numerically, are presented in section \ref{dos-section}. The theoretical basis which led to these analytical equations was published earlier \cite{ro-amg} and employs the Grassmann-field and generating functionals so that the density of states can be extracted from the single-fermion propagator. Section 5 finally contains the comparison with the Coulomb glass.

\section{The models under consideration, and their relationships}
\subsection{The spin glass models}
\label{models}
Let us begin with the SK-model, its exact fermionic representation (fermionized SK-model), and the generic fermionic SK-model, sometimes called ISG$_f$ of $f$SK-model.
These models agree for half-filling ($\mu=0$) and at $T=0$, and hence can be chosen for technical convenience.

We took advantage of the grand-canonical apparatus of generating functionals in Grassmann field theory, which has been elaborated for the fermionized version.
The fermionic SK-model, as the extreme localized limit of an itinerant spin glass model, contains certain aspects which can be compared with the Coulomb glass.


The SK-model together with its grand-canonical fermionic representation (fermionization) and its fermionic variant, listed by the spin Hamiltonian ${\cal H}_{SK}$ and the fermionic partners ${\cal H}_{fSK}$ respectively, are given by
\begin{eqnarray}
{\cal H}_{SK}&=&-\sum\limits_{i,j}J_{ij}S_i\, S_j-\sum\limits_{i=1}^N H_i\, S_i\quad\quad{\rm with}\quad S_i=\pm1\\
{\cal H}_{fSK}&=&-\sum\limits_{i,j}J_{ij}\sigma_i\sigma_j-\sum\limits_{i=1}^N (H_i\,\sigma_i-\mu\, n_i)\quad\quad{\rm with}\quad \sigma_i=n_{_{i\uparrow}}-n_{_{i\downarrow}}
\end{eqnarray}
where $\sigma_i\equiv n_{i\uparrow}-n_{i\downarrow}$, $n_{i\lambda}\equiv a_{i\lambda}^{\dagger}a_{i\lambda}$
in terms of the fermionic operators obeying $\{a_{\alpha},a_{\gamma}^{\dagger}\}=\delta_{\alpha,\gamma}$. The spin interaction $J_{ij}$ is assumed to be range-free and Gaussian-distributed with zero mean (for simplicity) and variance $J^2/N$.

The chemical potential $\mu$, assumed as nonrandom and homogeneous in this case, is chosen for the grand-canonical model Hamiltonian (2) in two ways as given by
\begin{eqnarray}
(2a) &&\mu=i\frac{\pi}{2}T \quad \text{ fermionized SK-model}\nonumber\\
(2b) &&\mu =0 \quad \text{ half-filled SK-model}\nonumber
\end{eqnarray}

The ingenious trick introduced by Popov and Fedotov \cite{popov-fedotov}, which allows fermionization for $\mu=\frac{i}{2}\pi T$, has been used before in many of our preceding papers (including its generalization to higher spin quantum numbers). This mapping is exact in the sense that the thermodynamics of the original model and of its fermionized version are identical.

The 2nd model is also classical, since all operators commute - quantum time-dependence can only be observed when correlations of odd numbers of fermionic operators at different times are considered. The chemical potential signals that a grand-canonical ensemble is employed. For half-filling at $T=0$, realized by $\mu=0$, both models agree.

The distribution of internal magnetic fields $P(h)$ \cite{thomsenthorpe} is defined for the SK-model by
\begin{equation}
P(h)=\frac{1}{N}\sum_i\langle \delta(h-\sum_j J_{ij} S_j)\rangle
\end{equation}
while the fermionic density of states of the fermionic SK-model is defined via the disorder-averaged single fermion Greens function by
\begin{equation}
\rho(E)=\sum_{\alpha} \langle\delta(E-E_{\alpha})\rangle=
-\frac{1}{\pi}Im\,\langle\,G_{ii}^R(E)\,\rangle
\end{equation}
where $G^R$ represents for the retarded and real-space-local fermionic Greens function. The bracket refers to statistical- and disorder-average over the Gaussian-distributed $J_{ij}$.
Standard many body formalism tells us how to obtain $G^R$ from the real-time-ordered $T=0$ Greens function $G_{ij}(t-t')=-\langle T_t \{a_i(t) a_j^{\dagger}(t')\} \rangle$
Their (Lehmann) spectral weight representations
$G_{ii}^R(E)=\int\limits_{-\infty}^{\infty} du\, \frac{\rho(u)}{E+i\delta-u}$
and
$G_{ii}(E)=\int\limits_{-\infty}^{\infty} du\, \frac{\rho(u)}{E+i\,sign(E)\,\delta-u}$
yield either a symmetric imaginary part $\rho(E)$ as spectral weight or the antisymmetric function
$\rho^{(a)}(E)\equiv sign(E)\,\rho(E) =-\frac{1}{\pi}Im\{G(E)\}$ with
$\rho^{(a)}(E)=-\rho^{(a)}(-E)$.

We want to expand the antisymmetric function $\rho^{(a)}(E)$ in the complete orthonormal set of eigenfunctions of a 1D harmonic oscillator.
We shall describe below in detail how many states are necessary to describe with increasing accuracy the deviation of the numerically evaluated anti-symmetrized DOS $\rho^{(a)}(E)$ from the (functional form of the) first excited state of the harmonic oscillator, $\psi_1(x)$. We add one subsection in order to motivate the analogies between variables used in the DOS or $P(h)$ and in the oscillator models' wavefunctions.

Finally we shall compare the density of states of the fermionic SK-model with that of
the Coulomb glass.
Pioneering work by Davies, Lee, and Rice \cite{davies} and by Gr\"unewald, Pohlmann, and W\"urtz \cite{wuertz} provided strong numerical evidence for the connection between the
Efros-Shlovskii pseudogap and glassy order in Coulomb glasses, which intend to model the deep localized regime of disordered electrons interacting by Coulomb interaction.

In these models, random chemical potentials were either modeled by a constant probability distribution on a finite width, while later on, M\"uller and Pankov \cite{mp} used a Gaussian distribution to obtain an effective pseudo-spin model and analyzed the ordered phase in great detail \cite{mp}.

Our objective is to display both the similarities and the differences between density of states of the fermionic SK-model and of the Coulomb glass. We report a new calculation on the replica symmetric approximation including the unstable temperature-regime and $T=0$. We are then in a position to appreciate the different types of pseudogaps created by RSB. For 3D M\"uller and Pankov found that the DOS vanishes quadratically at the pseudogap-center, while in the SK-case it behaves linearly for all dimensions.

\subsection{The oscillator models}
The one-dimensional harmonic oscillator is one of the most elementary models of quantum mechanics. Its eigenfunctions are usually written in terms of Hermite polynomials and Gaussian functions (while the 2nd linear independent solution of the differential equation, which is a confluent hypergeometric function, is excluded by boundary conditions). Using this complete orthonormal set of basis functions for representations of other models' solutions, does not imply a priori any physical relationship with the harmonic oscillator. Hence in general one cannot expect to gain much from such a representation.

The apparent resemblance between $P(h)$ of the $T=0$ SK-model (and hence of the density of states $\rho(E)$ of its half-filled fermionic model extension \cite{perez}) on one hand, and the modulus of the first excited state wavefunction $|\psi_1(x)|$ of a harmonic oscillator on the other hand, motivated us to look for the number of basis states needed to describe well the deviations. Another faint suspicion was feeded by the replica-symmetric spin glass solutions at $T=0$, which has a split Gaussian shape and reminded (partially) of the oscillator ground state.

Analyzing spin glass data in terms of such an expansion, we first found that less
basis functions are needed than expected.
Furthermore, for an excellent approximation (see below) it turned out that an
unconventional shift was needed. This model can be described as
\begin{equation}
\label{osc-model}
{\cal H}_{osc}=\frac{\hat{p}^2}{2m}+\frac12 m\omega^2
\left(|x|- x_0\right)^2+c_3 |x|^3+c_5\,|x|^5+...
\end{equation}
where $\hat{p}\equiv -i\hbar \frac{d}{dx}$. The introduction of the nonanalytic shift
helped to select only the first excited state in order to match accurately the DOS-data by only the first excited state. For simplicity we set again $\hbar=1$ and $m=1$.

The energy of the oscillator in the $n$-th excited state, given by
$$E_{osc}^{(n)}=\int\, dx\, \psi_n^*(x){\cal H}_{osc}\,\psi_n(x),$$
while the SK-energy $E_{SK}$ at $T=0$ will then be given by
\begin{equation}
\label{SK-energy}
E_{SK}=F_{SK}(T=0)=\int\limits_{-\infty}^{0} dh\,h\, P(h)
=\int\limits_{-\infty}^{\mu=0} dE\,E\,\rho(E)
\end{equation}
and using the results for $P(h)$ or $\rho(E)$ we confirm the recently derived high-precision value of the SK-energy at $T=0$ correctly up to $O(10^{-8})$.

Anticipating the result let us represent the SK-energy $E_{SK}$ as an integral over the quantum oscillator wavefunction by
\begin{equation}
E_{SK}=\int\limits_{-\infty}^0 dx\, x\,|\psi_1(x)|
\end{equation}
which is the direct translation of Eq.(\ref{SK-energy}).

The SK-energy $E_{SK}$ has little to do with the oscillator energy in the first excited state mentioned above. We do not introduce the special quantum oscillator as a simpler 'replacement'-model for the SK spin glass, but for the time being rather focus on the mere demonstration that its first excited state reproduces the functional form of $P(h)$ or $\rho(E)$, which are both two important quantities of the spin glass order at $T=0$.

\subsection{Distinguishing differential equations of the $T=0$
SK-model from Burgers equation}
In order to motivate the corresponding sets of variables needed in the functional mapping of DOS, $P(h)$ and one hand and the oscillator eigenfunctions on the other, we wish to return to the differential equations of the spin glasses at zero temperature.

It was observed that the Parisi scheme led to a recursive relation for the so-called exponent-correction-function, denoted by $expC$, in arbitrary finite order of RSB (discrete Parisi scheme). The recursion relation turned into a partial differential equation \cite{mjs-ro-pre,diss_mjs} in the so-called continuum limit.

Here, for the purpose of comparing it with a Burgers equation, we apply a different simple transformation.
It is sufficient to start from the original equation as given in Ref.\cite{mjs-ro-pre}) for fields $0<h\leq\infty$ (see also Ref. \cite{diss_mjs} which includes $h=0$)
\begin{equation}
\frac{-1}{q'(a)}\partial_a expC(a,h)+\frac12 \left[\left(\partial_h^2+2\,a\,\frac{h}{|h|}\partial_h\right)expC(a,h)
+a\left(\partial_h expC(a,h)\right)^2\right]=0,
\end{equation}
where the order function $q(a)$ is known to be monotonic and its derivative $q'(a)$ vanishes only in the limit $a\rightarrow\infty$.
Applying the transformation
\begin{equation}
\label{Yeqofa}
Y(a,h):=expC(a,h)+|h|-\frac12 \int\limits_0^a d\tilde{a}\,\tilde{a}\,q'(\tilde{a})
\end{equation}
simplifies further the PDE.
We had given a numerically satisfying analytic model function $q(a)$ for $T=0$ in Ref.\cite{ro-mjs-pre}, which differs everywhere within $0\leq a\leq\infty$ by only less than $O(10^{-3})$ from the exact solution.
This function is monotonic and we may switch to the unique inverse function $a(q)$ and, even better, choose $\tau\equiv 1-q$ as a new independent variable, in order to remove the inconvenient sign. We call this second variable pseudo-time, since, by means of the transformation (\ref{Yeqofa}), the PDE now matches a diffusion equation with diffusion constant
$\frac12$ and one nonlinear term.
Apart from the inverse order function $a(\tau)$ it resembles a Burgers equation as
discussed just below.
\begin{equation}
\label{Y-pde}
\frac{\partial Y(\tau,h)}{\partial\,\tau}=\frac12 \left[\frac{\partial^2}{\partial h^2} \, Y(\tau,h)+a(\tau)\,\left(\frac{\partial\, Y(\tau,h)}{\partial h}\right)^2\right]\quad,\quad \tau\equiv 1-q
\end{equation}
Apparently we obtain a modified Burgers equation with a coefficient $a(\tau)$ which diverges for small argument like the inverse square root for $q\rightarrow 1$, hence $a(\tau)\sim \tau^{-1/2}$.
In recent papers we concluded that $\{a=0,T=0\}$ and $\{a=\infty,T=0\}$ are too different critical points of RSB hosted by the SK-model. The latter limit belongs to $q\rightarrow1$, while the small-$a$ limit corresponds to $q\sim a\rightarrow0$ and $a(\tau=1)=0$.

In order to complete our discussion above, we recall that the Burgers equation $\partial_t u(t,x)=D\, \partial_x^2 u(t,x)-u(t,x)\partial_x u(t,x)$ turns into
$$\partial_t\, \phi(t,x)=D\,\partial^2_x\,\phi(t,x)-\lambda\,\left(\partial_x\phi(t,x)\right)^2$$
by means of $u=\partial_x \phi$ and $\partial_t\partial_x \phi=\partial_x\partial_t\phi$.

It is well-known that the Burgers equation has a redundant nonlinearity, and that it can be transformed into a diffusion equation by means of the Cole-Hopf transformation \cite{debnath}.

In the differential equation for the SK-model, Eq.(\ref{Y-pde}), however, the coupling function $a(\tau)$ of the nonlinear term prevents a linearization by means of the Cole-Hopf transformation.
This may indicate that an exact analytical solution of the SK-model at $T=0$ is much harder than the exact KPZ-solution in one dimension\cite{calabrese,sasamoto}. This argument does not yet refer to the particular initial condition of the SK-model at $T=0$.

\subsection{Randomly stirred SK-model differential equation versus KPZ-equation}
It is natural to consider also a (non-thermal) noise perturbation of the $T=0$ SK differential equation. In case of the pure Burgers equation this perturbation is called random stirring and results in the KPZ-equation, which describes random growth of a surface height above a substrate during particle deposition.

Let us perturb the SK-energy by a non-thermal $\delta$-correlated noise function $\eta(t,h)$ such that the characteristic differential equation (\ref{Y-pde}) takes a form comparable with the KPZ-equation
\begin{eqnarray}
&&\frac{\partial u(t,h)}{\partial t}=D \,\nabla_x^2 \, u(t,x)+\lambda\left(\nabla_x u(t,x)\right)^2+\eta(t,x)\\
&&\frac{\partial Y(\tau,h)}{\partial\,\tau}=\frac12 \left[\frac{\partial^2}{\partial h^2} \, Y(\tau,h)+a(\tau)\,\left(\frac{\partial\, Y(\tau,h)}{\partial h}\right)^2\right]+\eta(\tau,h)
\end{eqnarray}
The renormalization of the one-dimensional KPZ-equation yields the critical exponents exactly \cite{barabasi}
and it is known that simple rescaling did not lead to those exponents, except for the Edwards-Wilkinson behaviour near the trivial fixed point. This fixed point is however unstable for one space dimension. A simple power counting analysis which compared different terms of the differential equation had no access to the behavior near the stable finite fixed point for the effective coupling \cite{barabasi}.
An exact solution of the 1+1 dimensional KPZ-equation with flat initial conditions was
presented by Calabrese and Le Doussal \cite{calabrese}.

Let us reconsider the question whether the modified PDE, which describes the SK-model with randomly perturbed energy, may however allow power counting to be successful.
Its disadvantage, which is the divergent coefficient function $a(\tau)\sim 1/\sqrt{\tau}$ of the nonlinear term, may render this possible.

\subsubsection{Scaling regimes in the randomly-stirred SK-model differential equation}

We have a good analytical model for the $T=0$ order function $q(a)$ for all $a$ and
almost exact knowledge of the asymptotic behaviour close to the critical points at $a=0$
and $a=\infty$, as derived from a scaling theoretical \cite{ro-mjs-pre} explanation of high-precision numerical data \cite{mjs-ro-pre}. This can be used to estimate the role played by the different parts of the differential equation. The diffusive part dominates the small $a$-limit and the nonlinear term the large $a$ limit, respectively.

Under this assumption that the noise term does not essentially change the large-$a$ property $1-q\sim 1/a^2$ and hence the coefficient function $a(\tau)$,
we can estimate by power counting the competition between terms of the differential equation and, in analogy with the treatment of the KPZ-equation \cite{barabasi}, eventually obtain power laws and even critical exponents.

Rescaling the pseudotime by $\tau\rightarrow b^z \tau$, the analog of the real-space variable by $h\rightarrow b\, h$, and the field $Y$ by
$Y\rightarrow b^{\alpha} Y$ one gets for the small $\tau$-regime (large $a$-regime or $q(a)\sim 1-.41/a^2$ in terms of the spin glass order function $q(a)$)
$$b^{\alpha-z}\frac{\partial Y(\tau,h)}{\partial\,\tau}\cong b^{\alpha-2}\frac12 \frac{\partial^2}{\partial h^2} \, Y(\tau,h)+b^{2\alpha-z/2-2}
\frac{const}{\sqrt{\tau}}\,\left(\frac{\partial\, Y(\tau,h)}{\partial h}\right)^2
+b^{-(z+1)/2}\eta(\tau,h)$$
In the large $a$-regime or small $\tau$-regime the nonlinear term dominates over the diffusive one and one gets
\begin{equation}
\alpha=\frac34\, ,\quad z=1-\frac{\alpha}{2}=\frac52.\nonumber
\end{equation}
This simple power counting result does not replace a complete renormalization group calculation.
Note that in this language of dimensionless pseudotimes $\tau=1-q$ the above scaling behaviour refers to a critical short-time limit. It describes the behaviour of the function $Y(q)$ or $expC(q)$ for $q$-values close to the diagonal of the Parisi-matrix.

For the long $\tau$-regime, which refers to small $a$ in the original equation, one has $q(a)\sim a$ behaviour instead and the diffusive term dominates over the nonlinear one.
For this regime one is left with a randomly stirred diffusive equation in leading order and the behaviour is the equivalent of Edwards-Wilkinson (EW) behaviour in the KPZ growth-model for one space dimension (and one time dimension).
In this case one gets simply $z=2$ and $\alpha=\frac12$ as for the EW-limit of the KPZ-equation.

\section{Evaluation of the zero temperature density of states}
\label{dos-section}
We consider the spectral function of the fermionic SK model, which has the virtue of being mappable to the internal field distribution $P(h)$ of the standard SK-model \cite{perez}. Thus we can make use of the functional identity $\rho(x)=P(x)$.

The density of states function $\rho_{\kappa}(u)$ at a fixed order $\kappa$ of replica-symmetry breaking steps can be described by the following hierarchical integrals below.
They form the basis for the present numerical evaluation of $\rho_{\kappa}(E)$ for $\kappa=0,1,2,...,100$ and on a dense grid of energies $E$. An RSB solution for small orders $\kappa \leq 4$ can be found in \cite{ro-ds-prb}, which also considered arbitrary filling and an additional Hubbard interaction.

The following formula for the DOS holds for arbitrary order $\kappa$. Its evaluation allows to understand precisely the RSB-flow through the quasi-continuous regime at large $\kappa=O(10^2)$ towards $\kappa=\infty$.

\begin{equation}
 \rho_{\kappa}(E,\chi_1) = \frac{1}{2 \pi \sqrt{q_1 - q_2}} \exp(a_1 (|E|-\chi_1)) I_{\kappa}(|E|,0)\theta(|E|-\chi_1)
\end{equation}
which involves $\kappa$ nested integrals $I_n(E,h)$ given by the recursion relation
\begin{equation}
 I_{n}(E,h) = \int\limits_{-\infty}^{\infty} dh' \, g(h-h',q_{n+1} - q_{n+2}) \left(D_{n-1}(h')\right)^{\frac{a_{n+1}}{a_{n}} - 1} I_{n - 1}(E, h').
 \label{rho}
\end{equation}
The initial condition is given by the free-propagator like forms
\begin{equation}
 I_0(E,h) = \exp\left(-\frac{(E - \chi_1 - h)^2}{2(q_1 - q_2)}\right)
\end{equation}
with
\begin{equation}
g(x,y) = \frac{1}{\sqrt{2 y}} \exp\left(-\frac{1}{2} \frac{x^2}{y}\right).
\end{equation}
A second recursive structure is furthermore needed in Eqs.\ref{rho}, which now concerns the nested integrals of type $D_{n}(x)$. Their recursive structure is given by
\begin{equation}
 D_{n}(x) = \int\limits_{-\infty}^{\infty}dh\, \frac{\exp(\frac{(h-x)^2}{2(q_{n+1} - q_{n+2})})}{\sqrt{q_{n+1}-q_{n+2}}} \left( D_{n-1}(h) \right)^{\frac{a_{n+1}}{a_{n}}}
\end{equation}
with the initial condition
\begin{eqnarray}
 D_0(x) &=& \exp\left(-a_1 x + \frac{1}{2}a_1^2(q_1 - q_2)\right) \frac{1}{2}\left(1 - erf(\frac{x-a(q_1 - q_2)}{\sqrt{2(q_1 - q_2)}})\right)\nonumber \\
 &&\quad + \exp(a_1 x + \frac{1}{2}a_1^2(q_1 - q_2)) \frac{1}{2}\left(1 - erf(\frac{-x-a(q_1 - q_2)}{\sqrt{2(q_1 - q_2)}})\right).
\end{eqnarray}
The parameter sets $\{a_i\}$, $\{q_i\}$, and $\chi_1$, which are required in this
recursive set of equations, depend on the order $\kappa$.
They are inferred from a previous calculation, which reported their values \cite{mjs-ro-pre} for all $\kappa$ up to the maximum order $200$. They were obtained by extremization of the SK free-energy and their $T=0$ subset can be used since the fermionic SK- and the SK-model features coincide for $T=0$ and half-filling ($\mu = 0$), as mentioned in the model section \ref{models}. In particular the non-equilibrium susceptibility $\chi_1$ vanishes for $\kappa\rightarrow\infty$ \cite{mjs-ro-pre}, which gives rise to the perfect pseudogap.
\begin{figure}[here]
  \centering
  \resizebox{1.\textwidth}{!}{%
  \includegraphics{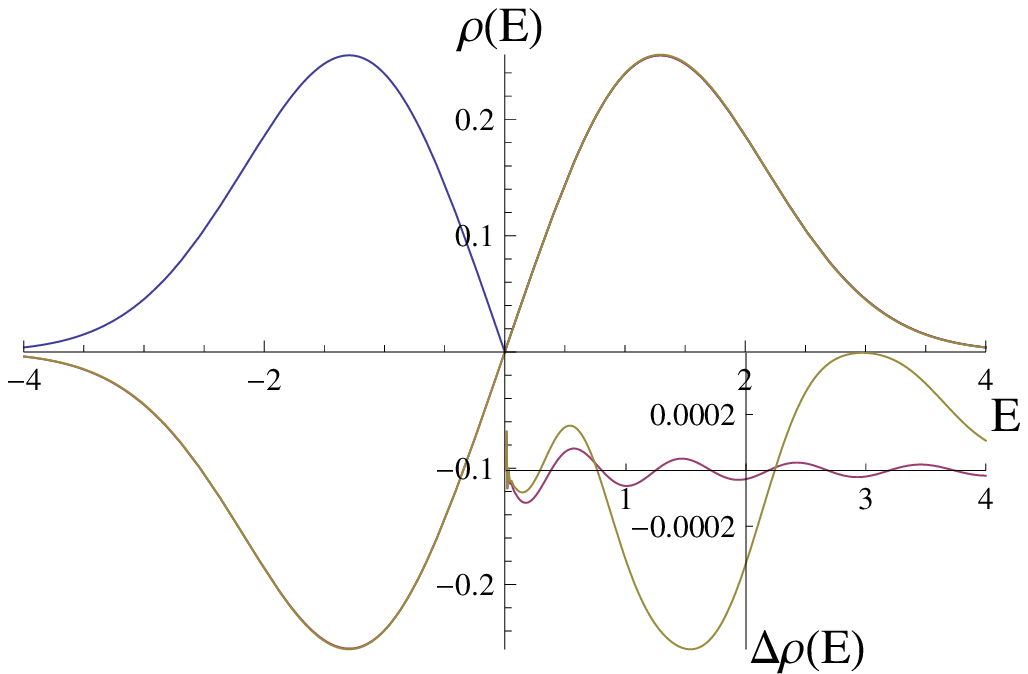}
   \hspace{-0.cm}\includegraphics{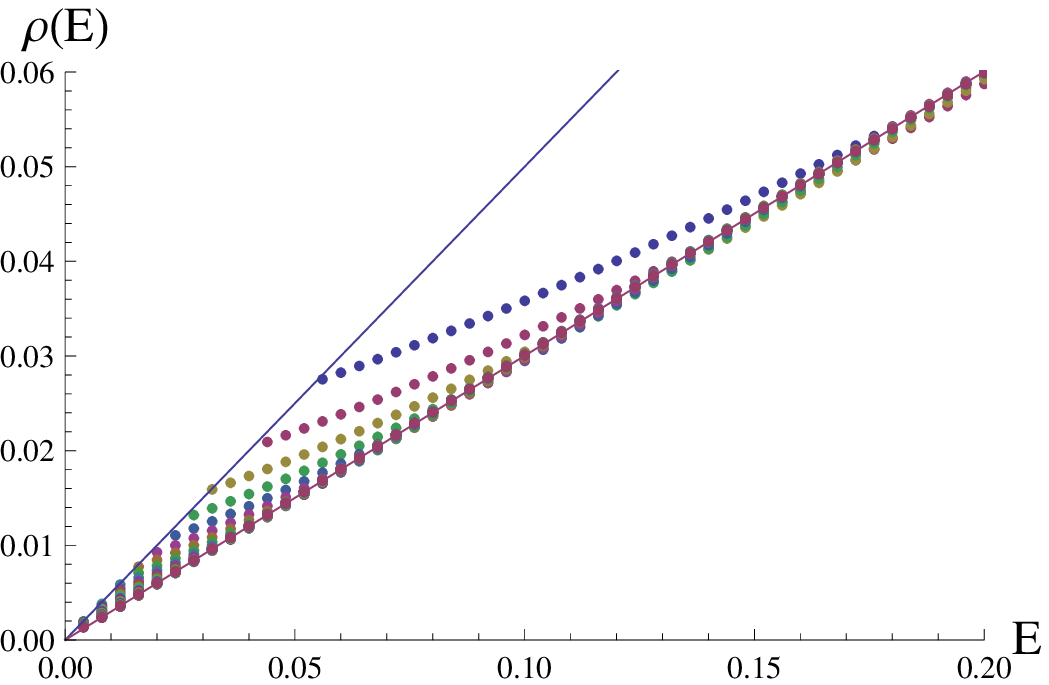}}
  \caption{(left figure:)
  The plot shows the numerical result $\rho_{100}(E)$ for the DOS at
$\kappa = 100$ (blue, symmetric curve) compared with its
partial sum $\sum_{n = 0}^{21} c_{100}^{(n)} \psi_{n}(E)$ of the
harmonic oscillator decomposition and the first excited state
$\tilde{\psi}_1(E)$ of the approximate Hamiltonian $H_{approx}$.
In the lower right corner the difference $\Delta \rho(E) = \rho_{100}(E) -
\rho_{approx}(E)$ is plotted for the two approximations.
Note that both approximations are correct up to
$\mathcal{O}(10^{-3})$ at each point.
\\
(right figure:) This example displays, for selected energies in the pseudogap regime, the flow of density of states data for the fermionic SK-Model under increased RSB-order $\kappa$. The approach of the fixed point DOS $\rho(E)\equiv lim_{\kappa\rightarrow\infty}\rho_{\kappa}(E)$ with slope $0.3$ \cite{thomsenthorpe} is obvious (even from a subset of the available $\kappa$).}
\end{figure}
\section{Analysis of the numerical high order RSB data}
Numerical results for the $T=0$ density of states have been evaluated for all $\kappa=1,2...,100$ orders of replica symmetry breaking.
One observes an incredibly fast convergence such that after $\kappa = 10$ any high $\kappa$ numeric solution can be effectively treated as the $\kappa = \infty$ result.
For the following calculations the data for $\kappa=100$ are used. The quality has been checked by comparing $\rho_{100}(E)$ with the fixed point function $\lim\limits_{\kappa\rightarrow\infty}\rho_{\kappa}(E)$ evaluated by using Pad\'e-series. The difference is negligibly small rendering either choice equally good.

Finding an analytic form of the density of states function is not only interesting but also useful for further calculations (e.g. Greens-Function for the model). In the course of finding a good fit for the data, it turned out to be quite practical to antisymmetrize the data by hand, thus getting rid of the non-analytical point at the origin.

Looking at this new data-set the similarity to the first excited state of the harmonic oscillator becomes rather striking. However the detailed analysis shows some necessary corrections as described in the next section.

\subsection{Calculating the overlap of $\rho(E)$ with harmonic oscillator eigenfunctions}

The eigenfunctions of the harmonic quantum oscillator, described by the Hamiltonian
$\hat{H_0} = \frac12 \hat{x}^2 + \frac12 \hat{p}^2$ in dimensionless variables, given by
\begin{equation*}
 \psi_n(x) = \frac{1}{\sqrt{\sqrt{\pi}n! 2^n}} H_n(x) \exp\left(-\frac{x^2}{2}\right)
\end{equation*}
with the Hermite polynomials $H_n(x)$, form the basis for our calculations.

Calculating the overlap coefficients
\begin{equation*}
 c_{\kappa}^{(n)} = \int dx \, \psi_n(x) \rho_{\kappa}^{(a)}(x)
\end{equation*}
quantifies the deviation of the function $\rho^{(a)}(x)$ from $\psi_1(x)$.

\subsection{Construction of an anharmonic Hamiltonian and functional mapping}
It turns out that it is possible to perturb the harmonic oscillator in a way that the DOS (here $\rho_{\kappa=100}^{\text{(a)}}$) coincides with the functional form of the oscillator's first excited state. The perturbation used in our case is rather unusual, since it involves the absolute values of the $\hat{x}$ operator, being $\vert\hat{x} \vert$.
Since the overlap-coefficients $c_{\kappa}^{(n)}$ go to zero rather fast and the computation of the change in all eigenfunctions due to the perturbation is impossible, we restrict ourselves to a subset of the first $22$ eigenfunctions. That renders an algebraic treatment of the Hamiltonian as a $22\times 22$ matrix possible.

We use the fact that eigenstates and eigenvalues of an anharmonic oscillator can be well approximated by diagonalization of the non-diagonal matrix representation $\langle \psi_k | H_{anh} | \psi_l \rangle$ in a subset of eigenstates $|\psi\rangle$ of the unperturbed harmonic oscillator.

\begin{eqnarray}
&&{H_0}_{mn} = (n + \frac{1}{2}) \delta_{mn} \\
&&\vert \hat{x} \vert_{mn} = \int dx \, \psi_m(x) \vert x \vert \psi_n(x) \\
&&\hat{H}_{p} = \hat{H}_0 + \alpha \vert \hat{x} \vert + \beta \vert \hat{x} \vert^3 + \gamma \vert \hat{x} \vert^5
\end{eqnarray}

In order to find the correct coefficients $\alpha, \beta, \gamma$ for the given perturbation it is useful to define the following function $l(\alpha,\beta,\gamma,\lambda)$, introducing a new eigenvalue $\lambda$.
In order to find the correct variational parameters solving $\hat{H}_p \vec{c}=\lambda\vec{c}$ it is useful to define the function
\begin{equation}
l(\alpha,\beta,\gamma,\lambda) = \Vert \hat{H}_p[\alpha,\beta,\gamma] \vec{c} - \lambda \,\vec{c} \Vert
\end{equation}
which controls the numerical accuracy. The function is constructed in a way such that the correct coefficients minimize it.
\subsubsection{Anharmonic quantum oscillator eigenstate modeling the spectral function}

The results for the leading parameters $\alpha,\,\beta,\,\gamma$ of the oscillator potential and for the energy eigenvalue of the first excited state are given by
\begin{equation}
\framebox{$\alpha = -0.899165,\quad \beta = -0.003268,\quad \gamma = 0.0000405,\quad \lambda = 0.364335$}
\end{equation}
and the square distance from the minimum is $l = 9.9\, 10^{-7}$.
The minimum was found by employing the Minimize-function implemented in Mathematica.

While the focus is on the first excited state, Fig.3 includes the three lowest states for the given potential parameters
\begin{figure}[here]
\label{fig:pot}
\centering
\resizebox{.7\textwidth}{!}{%
\includegraphics{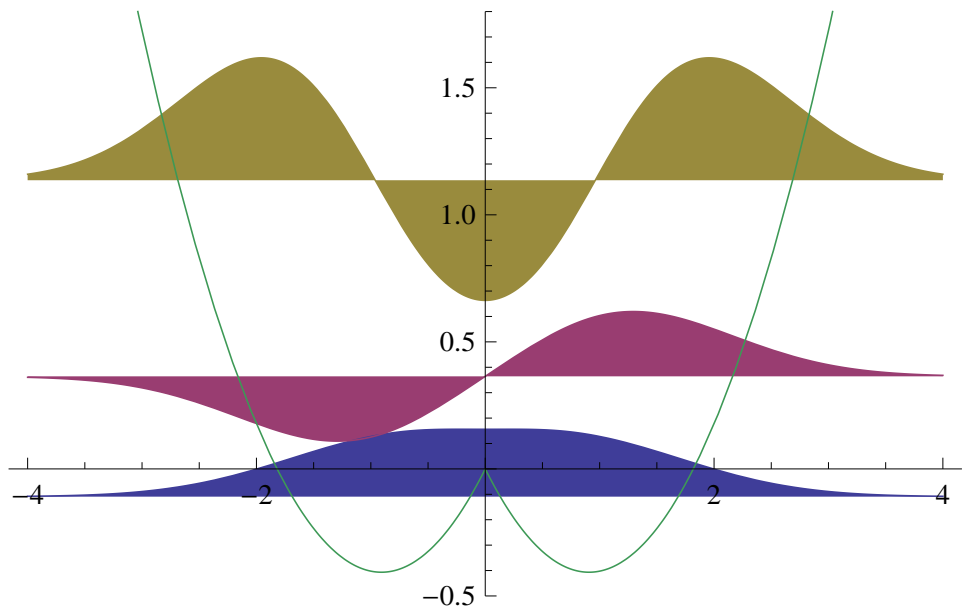}}
\caption{The three lowest eigenstates of the nonanalytic cubic potential $V(x)=\frac12 x^2+\alpha |x|+ \beta |x|^3+ \gamma |x|^5$ are displayed together. The position of their base-line indicates the corresponding energy eigenvalue.
The first excited eigenstate (odd parity) reproduces the density of states $\rho(x)sign(x)$ of the fermionic SK-model, and hence $P(h)sign(h)$ of the standard SK-model too, with full Parisi symmetry breaking at zero temperature.}
\end{figure}

Even the cubic term could be neglected and still the perturbed Hamiltonian produces a very good fit of the calculated DOS with slightly modified parameters
\begin{equation}
\framebox{$\alpha = -0.9225,\quad\beta=\gamma=0,\quad \lambda = 0.3444$}
\end{equation}
and $l = 7.3\, 10^{-6}$.

Thus, by keeping only linear and quadratic terms, one may rewrite the oscillator model
with an unconventional quadratic potential shifted by a constant, but using $\vert \hat{x} \vert$ instead of the normal space operator $\hat{x}$. Alternatively, one may use the standard real space operator together with an unconventional shift, which flips its sign at $x=0$.
The potential function as included in Fig.3 shows a double-well feature induced by the unusual shift; a difference between the cubic and the quadratic approximation is invisible on the given scale.

The resulting approximate Hamiltonian assumes the following form

\begin{equation}
 \hat{H}_{\text{approx}} = \frac{1}{2} \left[ \hat{p}^2 + (\vert \hat{x} \vert - x_0)^2 \right]=\,-\frac{1}{2}\partial_x^2 + \frac12(x - x_0\, sign(x))^2
\end{equation}

In order to control the finite cutoff imposed on the number of harmonic oscillator basis functions, the effective model with nonanalytic shift and small anharmonic terms, as derived above, can be plugged into a Schr\"odinger equation, which is then solved numerically.

Neglecting anharmonic terms of $O(|x|^3)$ because of their smallness but keeping only the nonanalytic shift (hence the $|x|$-term), the approximate model
Hamiltonian $\hat{H}_{\text{approx}}$ leads to the differential equation

\begin{equation}
 - \partial_x^2 \psi(x) + (x^2  + 2 \alpha \lvert x \rvert - 2 \lambda) \psi(x) = 0
\end{equation}

The numerical solution of this equation confirms the results of our large but finite matrix diagonalization approach.
This means, not only the desired first excited state $\psi_1$ is well approximated, but the other stationary states too. We used this as an independent control of our procedure.

\section{A comparison with the mean field solution of the Coulomb glass}
Since we focus in the present article on characterization and classification of replica symmetry breaking at lowest temperatures including $T=0$, we wish to compare its role in two different host models by means of pseudogap features.
There seems to be enough evidence that the host-model, which undergoes this type of symmetry breaking, can have an influence. While the universal critical behavior of replica symmetry breaking at $T=0$ is primarily caused by its hierarchical structure, it can yet belong to different universality classes depending on the host-model.

Recent work on the so-called Coulomb glass (CbGl)  offers an interesting comparison with the fermionic SK-type of RSB. Both models can be viewed to describe the insulating phase of a disordered itinerant fermionic systems - the source of glassiness being either a random chemical potential (CbGl) or a frustrated spin interaction respectively.

A selfconsistent structure was derived by M\"uller and Pankov for the Coulomb glass model \cite{mp}. Their starting point, was the transformation to an effective spin glass model, justifying a mean field description by means of the long-range nature of the bare Coulomb interaction.
The resulting model appeared as a type of SK-model with an effective $1/r$-dependent pseudo-spin interaction. This spatial dependence provokes a dimensional dependence.

An intriguing feature of the M\"uller-Pankov selfconsistent theory is the
large depletion of the fermionic density of states already far above the glass transition, identified by the Almeida-Thouless temperature.
The dip of $\rho(E)$ apparently becomes very pronounced already above $T_g\equiv T_{AT}$, and the authors have shown the changeover into the final Coulomb pseudogap by taking the full Parisi RSB into account. The similarity with the Efros-Shklovskii result was demonstrated. The evaluation close to $T=0$ indicated a quadratic decay of the DOS in 3D, in agreement with the ES-law. The authors also reported some results for the two-dimensional case.

\begin{figure}[here]
\resizebox{.45\textwidth}{!}{%
\includegraphics{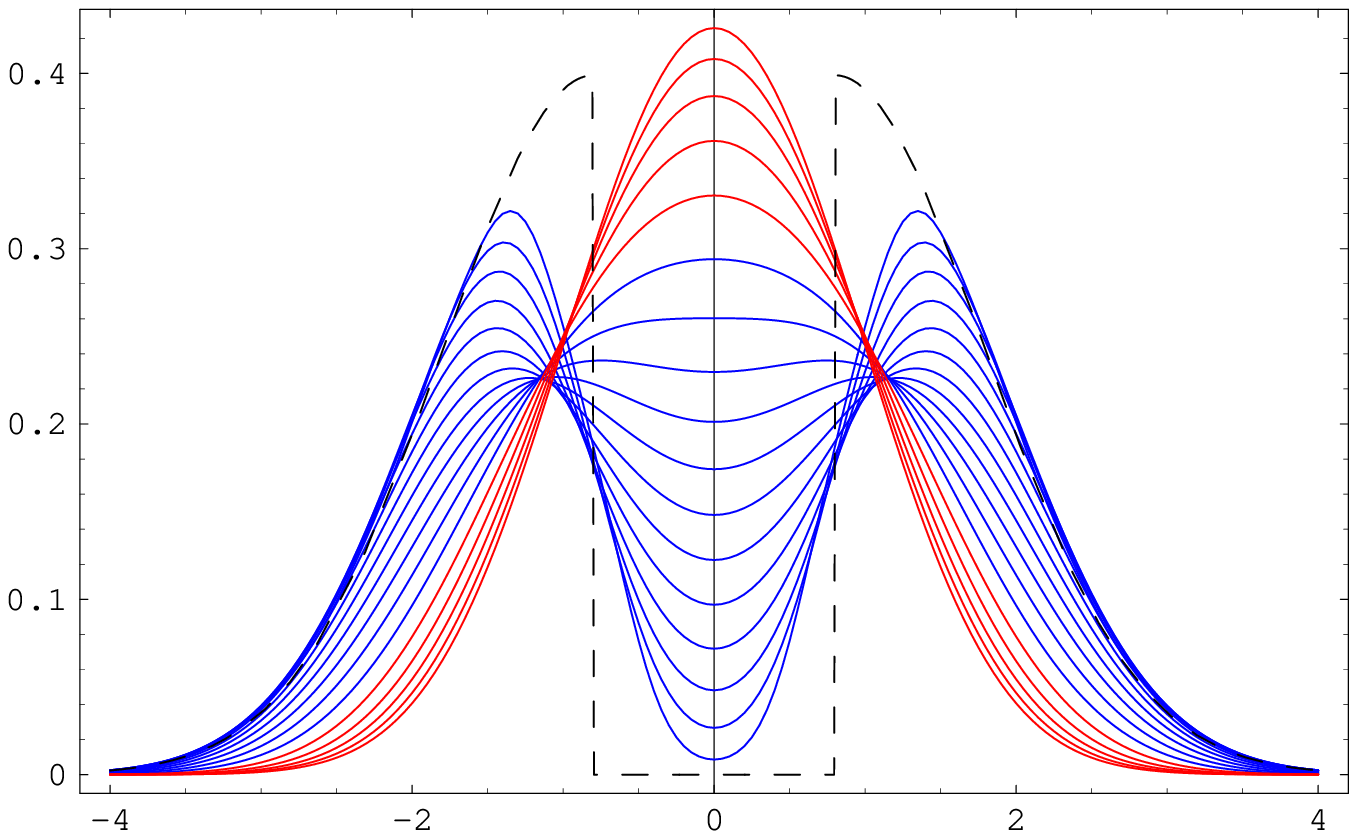}}
\resizebox{.45\textwidth}{!}{%
\includegraphics{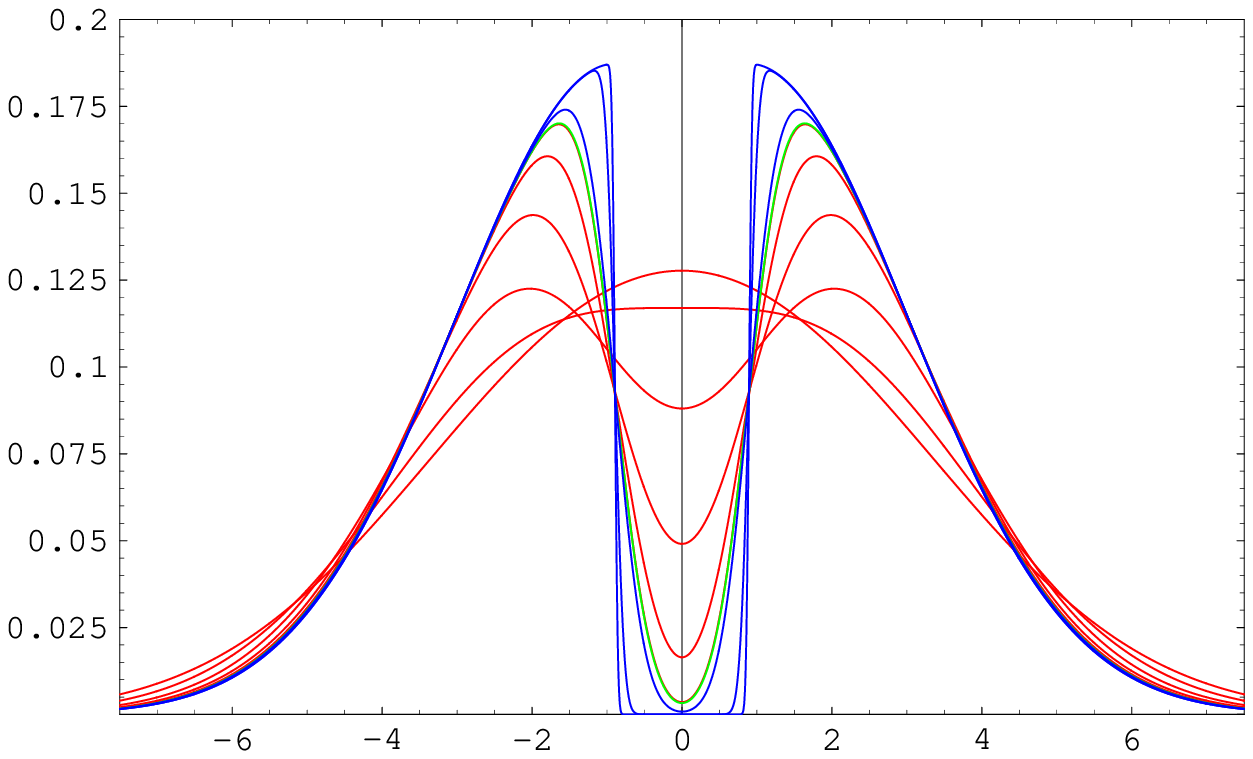}}
\caption{left panel shows density of states (DOS) of the fermionic SK-model for temperatures above $T_c$ (red) and below (blue), including the exact analytical result at $T=0$ (dashed-black) without replica symmetry breaking. Right panel: For comparison, the analogous replica-symmetric (RS) re-calculation on the basis of the M\"uller-Pankov Coulomb glass (CbGl) equations is shown. The depletion of the DOS almost looks like a pseudogap already at temperatures far above the Almeida-Thouless $T_g$ but, as the temperature decreases further towards zero, a hard gap is formed in the RS-approximation of the CbGl as in the fermionic SK-model. Displayed curves are shown for temperatures above $T_g$ (red), at $T_g$ (green), below (blue) down to $T=10^{-3}$.
The true pseudogaps (see above for the fermionic SK) and the Coulomb glass \cite{mp} obey different power laws and differ strongly from the shown RS-solutions.}
\label{RS-gap}
\end{figure}

In order to see the quantitative genuine effect of replica symmetry breaking in the CbGl-model, we evaluated their self-consistent coupled equations {\it without} replica symmetry breaking also below $T_g$. As initially suspected, the difference between solutions with broken and unbroken symmetry becomes larger with temperature decreasing towards zero. At first sight, the pseudogap seemed to be shaped already above $T_g$ (resembling a lot the curves of the fermionic SK-model in a certain range below its $T_c=T_g$, see Fig.\ref{RS-gap}), with only small corrections to be expected from symmetry breaking as $T$ decreases further. But we found that, within the replica-symmetric self-consistent structure, the depletion starts to form a hardgap, which, under certain assumptions, can even be elaborated analytically. The numerical results down to $T=10^{-3}$ and moderate disorder are shown in Fig.\ref{RS-gap}. Using the same notation and units as \cite{mp}, the hardgap half-width equals $h_0$, which is evaluated to be $h_0\cong 0.894306684$. This value could be given more accurately, but is is not needed here, it also shows that only parts of the calculation can be done numerically.

Thus, one might say that the pseudogap-like form, in part developed already above $T_g$, would be destroyed without RSB as $T\rightarrow 0$. Hence RSB is again responsible for the existence and shape of the pseudogap. This is one of the crucial points for a comparison with the fermionic SK-model: the replica symmetry breaking is fully responsible for the pseudogap in the low temperature limit.

The second point is that the CbGl-pseudogap shows, according to M\"uller and Pankov\cite{mp}, a $d$-dependent power law and in particular a quadratic decay near the center of the pseudogap for 3D, while the SK-model shows linear behavior (for all $d$). Thus the host of replica symmetry breaking can have an effect on its critical behavior, since, in our point of view, the small energy deviation from the gap-center belongs to the critical point $a=\infty$ of the order function $q(a)$ as derived recently in
Ref.\cite{ro-mjs-pre}, and hence to small pseudo-times $\tau=1-q$.

\section{Conclusions and remarks}
The main purpose of the present article is to show that the internal magnetic field distribution and the density of states as central quantities of SK spin glass physics
can be described by the functional form of the first excited state of a weakly anharmonic quantum oscillator with nonanalytic shift. This meant the harmonic potential to be of the form $(x-x_0\, sign(x))^2=(|x|-x_0)^2$ instead of $x^2$.

Puzzled by the unexpected relationship between spin glass and oscillator with a non-analytical shift, we searched for other singular anomalous oscillator models.
We became aware of recent work by Ritort on wedge-shaped oscillator potentials \cite{ritort}. He reported glassy behavior and aging effects in so-called  generalized oscillator model(s) GOM, including the case of a pure and positive $|x|$-potential.

In our independent study
the $|x|$-term has opposite sign and required (at least) $x^2$-term for stability. The nonanalytic linear term was required as an optimized shift to reproduce correctly the $\rho(E)$-data (and by analogy $P(h)$ as well) with a minimal number of oscillator base functions.

Let us also mention recent numerical work by Boettcher et al \cite{boettcher} on small- and moderate-sized spin glass models (including SK) for the magnetic field distribution at $T=0$.

The field distribution of the 3D Coulomb glass, and also those of XY- or Heisenberg spin glass models, show rather quadratic small field behavior. This raises the question whether they can be represented by oscillators too, but not by one single excited state. For continuous dimensions $D$ one may also wonder how a $\rho(E)\sim (E-E_F)^{D-1}$ can be obtained. Perhaps fractional derivatives can generalize the differential equations such that these power laws can be represented by oscillator models. These questions appear quite open and might lead to more insight into the spin glass oscillator relationship.

\section{Acknowledgements}
We wish to thank the DFG for support under Op28/7-2. One of us is indebted to A. Crisanti, C. De Dominicis, and T. Sarlat for useful remarks, and for hospitality extended to one of us (R.O.) at the CEA Saclay, where part of this work was initiated.
We are also indebted to David Sherrington for his long-term interest in our research on low $T$ and $T=0$-RSB, and for his constant emphasis on the importance of $P(h)$.

\end{document}